\documentstyle[aps,prl,epsf]{revtex}

\begin{document}
\twocolumn[

\hsize \textwidth\columnwidth\hsize
\csname @twocolumnfalse\endcsname
\setcounter{equation}{0}

\title{Do columnar defects produce bulk pinning?}

\author{M.V. Indenbom,$^{1,2}$ C.J. van der Beek,$^{2}$ M. 
Konczykowski,$^{2}$ and F. Holtzberg$^{3}$}
\address{$^{1}$Institute for Solid State Physics R.A.S., 142432 Chernogolovka,
Moscow district, Russia \\
$^{2}$Laboratoire des Solides Irradi\'{e}s, Ecole Polytechnique, 91128
Palaiseau, France \\
$^{3}$Emeritus, IBM Thomas J. Watson Research Center, Yorktown Heights, N.Y. 
10598, U.S.A.}

\address{}

\date{\today}

\maketitle

\begin{abstract}
From magneto-optical imaging performed on heavy-ion
irradiated YBa$_{2}$Cu$_{3}$O$_{7-\delta}$ single crystals, it is
found that at fields and temperatures where strong single vortex pinning by
individual irradiation-induced amorphous columnar defects is to be expected,
vortex motion is limited by the
nucleation of vortex kinks at the specimen surface rather than by
half-loop nucleation in the bulk. In the material bulk, vortex motion
occurs through (easy) kink sliding. Depinning in the bulk
determines the screening current only at fields comparable to or larger
than the matching field, at which the majority of moving vortices is not
trapped by an ion track.

\end{abstract}

] 
\narrowtext

\begin{figure}[t]
	\vspace{-1mm}
\centerline{\epsfxsize 3.5cm \epsfbox{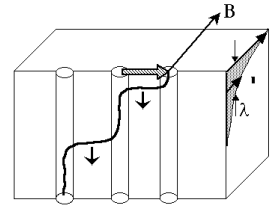}}
\caption{ \label{Surface}
Surface depinning of a vortex (bold line) from
the columnar defects (cylinders). Short arrows indicate the vortex
kink sliding down from the surface, producing a vortex drift to
the right. The surface critical current distribution in
the $\lambda $--layer is sketched on the right hand ``crystal face".
}
\end{figure}

Columnar defects created by heavy ion irradiation provide
very efficient vortex pinning in high temperature superconductors
\cite{First}. Nevertheless, because the column radii are very
homogeneous over their length\cite{Atomic}, it is not clear how 
the motion of even slightly misaligned vortices be can inhibited.
Aside from the case where the angle
between the applied magnetic field and the ion tracks 
is deliberately chosen to be non-zero \cite{Schuster}, misalignment
between vortices and columns arises
from the presence of the shielding current itself, since the
latter implies not only a gradient of the vortex density but also vortex
line curvature. The problem is illustrated in Fig.~\ref{Surface}.
If the vortex lines are inclined with respect to the ion tracks,
vortex kinks connecting segments trapped
by the columns can easily slide along them. The force opposing
this motion is determined by the background pinning by point defects.
Hence, the critical current will be orders of magnitude lower
than that corresponding to the depinning of vortices from the columns by
a (double) kink nucleation process\cite{Nelson92}. The large observed
critical currents\cite{First}, as well as the moderate anisotropy for
vortex motion within and across the plane containing the irradiation direction
and the $c$-axis in obliquely irradiated DyBa$_{2}$Cu$_{3}$O$_{7-\delta}$
single crystals\cite{Schuster}, indicates that kink sliding cannot be the
main mechanism
limiting flux motion type-II superconductors with correlated
disorder. Rather, it was suggested\cite{Schuster} that,
in crystals of thickness $d$ much greater than the penetration depth
$\lambda$, it is the nucleation of vortex kinks at the crystal surface that
plays
this role (shaded arrow in Fig.~\ref{Surface}). By consequence, the
critical current
only flows in a surface layer of thickness $\sim \lambda$; kink sliding
causes the current
density $j(z)$ in the bulk to drop to a value that is too
small to induce vortex--kink or half--loop nucleation.

In this paper, it is verified that vortex motion in irradiated
YBa$_{2}$Cu$_{3}$O$_{7-\delta}$ (YBCO) single crystals indeed proceeds
through the
``hard'' nucleation of kinks at the surface followed by ``easy'' kink sliding
into the crystal bulk, {\em irrespective of the
relative alignment between vortex lines and ion tracks}. Our method
relies on the measurement of the thickness dependence of the crystals'
self-field: if the critical current
flows only within a surface layer, the integrated shielding current
$J = \int_{-d/2}^{d/2} {j(z)dz}$, and hence the hysteretic parts of the
magnetic moment and of the induction measured at the crystal surface,
should be independent of the thickness.

The most reliable way to demonstrate a thickness (in)dependence
of the self--field, excluding the usual scatter of the crystal properties,
is to observe the flux penetration into a flat sample with big surface steps.
Supposing that the bulk current $j$ is homogeneous,
the characteristic field for penetration of perpendicular flux into
a flat superconducting plate is proportional to $J = jd$\cite{Flat};
a much easier flux penetration into the thinner parts of
such superconducting samples has clearly been observed using
magneto-optics, and was reproduced in model calculations\cite{Inhomo,Step}.
For surface-like pinning, in which only a surface current $J_{s}$
is present, $J = 2J_{s}$ and flux penetration should be like that into
a crystal of constant thickness.

\begin{figure}[t]
\centerline{\epsfxsize 7cm \epsfbox{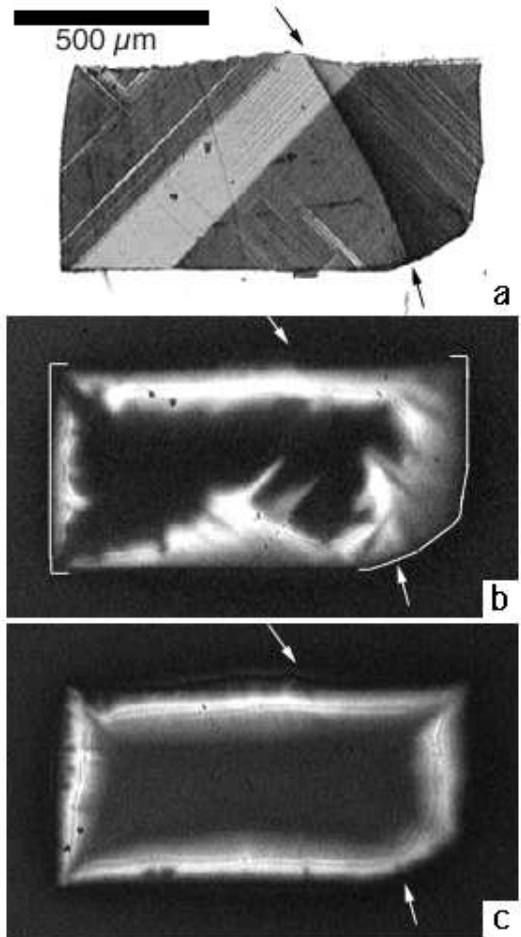}}
\vspace{0mm}
\caption{ \label{BeforeAfter}
Regularization of flux penetration by heavy ion irradiation:
(a) Reflected polarized light photograph of the surface
of a YBCO single crystal. The crystal has a large surface step 
(indicated by arrows), dividing it
into RHS and LHS parts of thickness 10 and 20 $\mu$m, respectively.
(b) and (c) show the remanent induction on the crystal surface after the
application and removal of a field $H_{a} =$ 360~G $\parallel
c \parallel$ ion tracks:
(b) before irradiation, $T = 40$ K; (c) after irradiation
with 6 GeV Pb ions, $T = 80$ K. The arrows indicate the position 
of the step.
}
\end{figure}

YBCO single crystals were grown in Au crucibles and annealed in oxygen in Pt 
tubes as described elsewhere \cite{Growth}. For our experiments we have 
selected crystals with pronounced as--grown surface steps, in order to have a thickness
variation of at least a factor 2 over the crystal length. Microscopic observations in 
reflected polarized light revealed all crystals to be twinned 
(Figs.~\ref{BeforeAfter}(a) and 

\begin{figure}[t]
\centerline{\epsfxsize 8.5cm \epsfbox{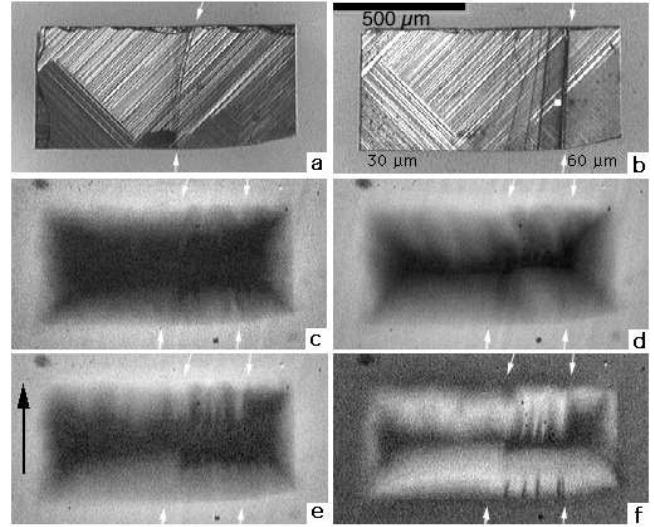}}
\caption{ \label{Half}
Role of surface steps in the flux penetration into an YBCO crystal
with columnar defects:
(a) Top surface of a crystal with a $15\mu$m step crossing it in the
center (at the arrows); (b) Mirror image of the bottom surface, 
with one large 10~$\mu$m step at the very right (see arrows) 
and a number of smaller steps. The crystal
thickness monotonically increases from left to right.
(c,d) Homogeneous flux penetration into the ZFC
crystal ($T = 85$ K).
(c) $H_{a} =$ 177~G, (d) $H_{a} =$ 359~G ($\parallel c \parallel$ ion
tracks);
(e,f) Image of the perpendicular induction on the crystal top surface,
after cooling to 85 K in a constant in-plane field $H_{\parallel} =$ 80~G applied in 
the direction of the bold arrow, and the subsequent application of  
$H_{\perp} =$ 169~G (e), and (f) a reduction of $H_{\perp}$ to 84~G after 
application of 253~G.
}

\end{figure}

\noindent 
\ref{Half}(a,b)). The crystals were irradiated at GANIL in Caen,
France, using a beam of
6 GeV Pb ions oriented parallel to the crystalline $c$-axis.
The track density $n_{d} = 5\times10^{10} {\rm cm^{-2}}$ corresponds
to the irradiation dose, and to a matching field $B_{\phi}= \Phi_{0}n_{d} =
10$ kG ($\Phi_{0}$ is the flux quantum). Flux penetration before and after the
irradiation was studied by means of the magneto-optical imaging technique
using ferrimagnetic garnet indicators with in-plane anisotropy\cite{MagnOpt}.
On all images of the flux distribution presented in this paper the higher
value of the image intensity corresponds to the higher value of the local
induction.

In Fig.~\ref{BeforeAfter} we present images of the flux penetration into 
one of the crystals before and after the irradiation. This  crystal has 
one large surface step, separating  it into two parts of thickness
10~$\mu$m and 20~$\mu$m respectively.

\noindent 
Fig.~\ref{BeforeAfter}(b) shows the remanent induction before the
irradiation, after the application and removal of an  applied field
$H_{a} = 360$ G $\parallel c$ at $T = 40$ K. Owing to the
crystal's twin structure,
flux penetration before the irradiation is rather irregular. This irregularity
was observed  in  all other crystals. 
Nevertheless, flux penetration into the thin right hand side (denoted 
in the Figure by the white bracket) is
clearly easier than that into the thick left hand part (also with 
bracket) of the crystal, which has a similar twin structure.

The introduction of columnar defects drastically changes the flux
penetration pattern. Because of the very substantial increase in
shielding current, the temperature had to be increased to 80 K in order
to observe penetration over a distance comparable to that before irradiation
(Fig.~\ref{BeforeAfter}(c)). Pinning by columnar defects
is seen to dominate all other pinning: if any influence of
the twin boundaries on the flux penetration is present,
it can no longer be discerned\cite{Schuster}. More important, flux now
penetrates {\em equally far into the thick and thin parts of the crystal}
in accordance with the hypothesis of surface depinning.

This finding is corroborated by measurements on a specially prepared
rectangular sample, cut from another  irradiated crystal in order to 
have a series of surface steps  of the same sign, oriented 
perpendicularly to its longer

\noindent  sides. Fig.~\ref{Half}(a,b) shows that this sample
has a big step of height $\sim 15$ $\mu$m on the top surface, dividing it into
two roughly equal parts of thickness 30~$\mu$m and 50-60~$\mu$m respectively.
On the bottom surface (the mirror image of which is shown in Fig.~\ref{Half}(b)
for easier comparison with Fig.~\ref{Half}(a)), there is another
large step of height $\sim 10$ $\mu$m, together with a number of smaller steps
of height $\sim 1$ $\mu$m.  The twin patterns revealed in reflected
polarized light are equivalent on both crystal sides, and are not
interrupted by the steps, thus showing the perfect continuity of the
sample. Subsequent magneto-optical imaging of the
flux distribution was carried out on the top surface. Again applying a
field parallel to the columnar defects, {\em i.e.} perpendicular to the plane of the
zero--field cooled sample, we observed the same striking phenomenon: the flux
penetration pattern appears as if the crystal had constant thickness (Fig. \ref{Half}(c,d)).
The distance over which flux penetrates is the same along all the sample
edges, {\em i.e. $J$ is thickness independent}. Note that the small irregularities
in flux penetration at the upper edge in Fig.~\ref{Half}(d) may be ascribed
to the defects caused by cutting (seen in Fig.~\ref{Half}(a)).

\begin{figure}[]
	\vspace{-2mm}
\centerline{\epsfxsize 8cm \epsfbox{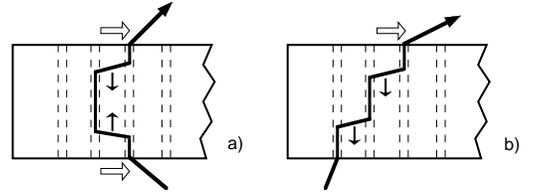}}
\vspace{1mm}
\caption{ \label{Model}
Vortex kink motion in perpendicular (a) and inclined field (b):
(a) Kinks nucleate at both crystal surfaces, slide into the interior and
anihilate there with each other -- the critical current of kink nucleation
flows on the both surfaces;
(b) Unipolar kinks nucleate at the vortex ``leading head'' and move down to
the opposite crystal surface -- the critical current flows on the upper
surface only.
}
\end{figure}

The above result constitutes strong evidence in
favor of the model in which depinning of vortices from parallel columnar
defects is limited by the nucleation of vortex kinks at both crystal surfaces,
the critical current being the surface current necessary for this
process (Fig.~\ref{Model}(a)). Vortex depinning in the situation
where the field is applied parallel to the columns thus
resembles depinning in the case where either are misaligned\cite{Schuster}.
Simultaneously, it is a well--known fact that the  magnetic moment of
heavy--ion
irradiated YBCO rapidly decreases when the angle between the applied field and
the columns is increased\cite{First}. It is therefore interesting to learn how
tilting the field affects surface depinning. For this, the same crystal was
cooled in a field $H_{\parallel} =$ 80~G directed parallel to the
crystal plane and parallel to its shorter sides (as indicated in Fig.
\ref{Half}(e)). The in-plane field was not changed during the subsequent
application of a perpendicular field $H_{\perp}$. Although the penetration
of $H_{\perp}$ appeared to be somewhat more pronounced in the thinner
left hand part of the crystal
(Fig. \ref{Half}(e)), the difference in penetration depth between the
two parts was considerably less than what should be expected
for bulk pinning, would this have been relevant after relieving the
vortex confinement to the columnar defects.

We also observe an intriguing easy flux motion along the small
surface steps on the bottom face of the crystal. Such a pronounced
influence of these steps is not to be expected in case bulk pinning is dominant.
The perpendicular induction (directed towards the observer) clearly
penetrated further along the steps at the upper edge (Fig.~\ref{Half}(e)). 
When the applied field was reduced, flux left the crystal preferentially at the lower
edge, while the features arising from the earlier preferential
penetration at the upper edge remained frozen (Fig. \ref{Half}(f)).
Reversing

\noindent the sign of either $H_{\perp}$ or $H_{\parallel}$
reversed the sense of easy flux motion along the steps:
flux now penetrated preferentially at the lower edge in increasing
$H_{\perp}$ and at the upper edge in decreasing $H_{\perp}$.

The motion of inclined vortices is mediated by {\em unidirectional}
kink sliding from the surface with
leading vortex end, where kinks nucleate, to the opposite surface
(see Fig.~\ref{Model} (b)). The observed easy
flux penetration along the sharp small steps on the bottom surface of the
crystal
is a result of easy kink nucleation at these steps.
The big step on the top face is smooth and does not affect
surface kink nucleation. The need to nucleate the kinks on one surface only
restricts the critical current flow to this surface (cmf. Fig.~\ref{Model}), which
explains the fact
that inclined vortices penetrate the crystal approximately twice as far
for the same temperature and $H_{\perp}$, as
well as the rapid decrease of the sample magnetic moment when the applied
field
is tilted from the track direction. The strong sensitivity to surface defects
is another fact supporting the idea of surface-like pinning:
as in the case of the Bean-Livingston surface barrier\cite{Livingston64} the
nucleation of vortex kinks is considerably
facilitated by small but sharp surface irregularities.

\begin{figure}[]
\centerline{\epsfxsize 8.5cm \epsfbox{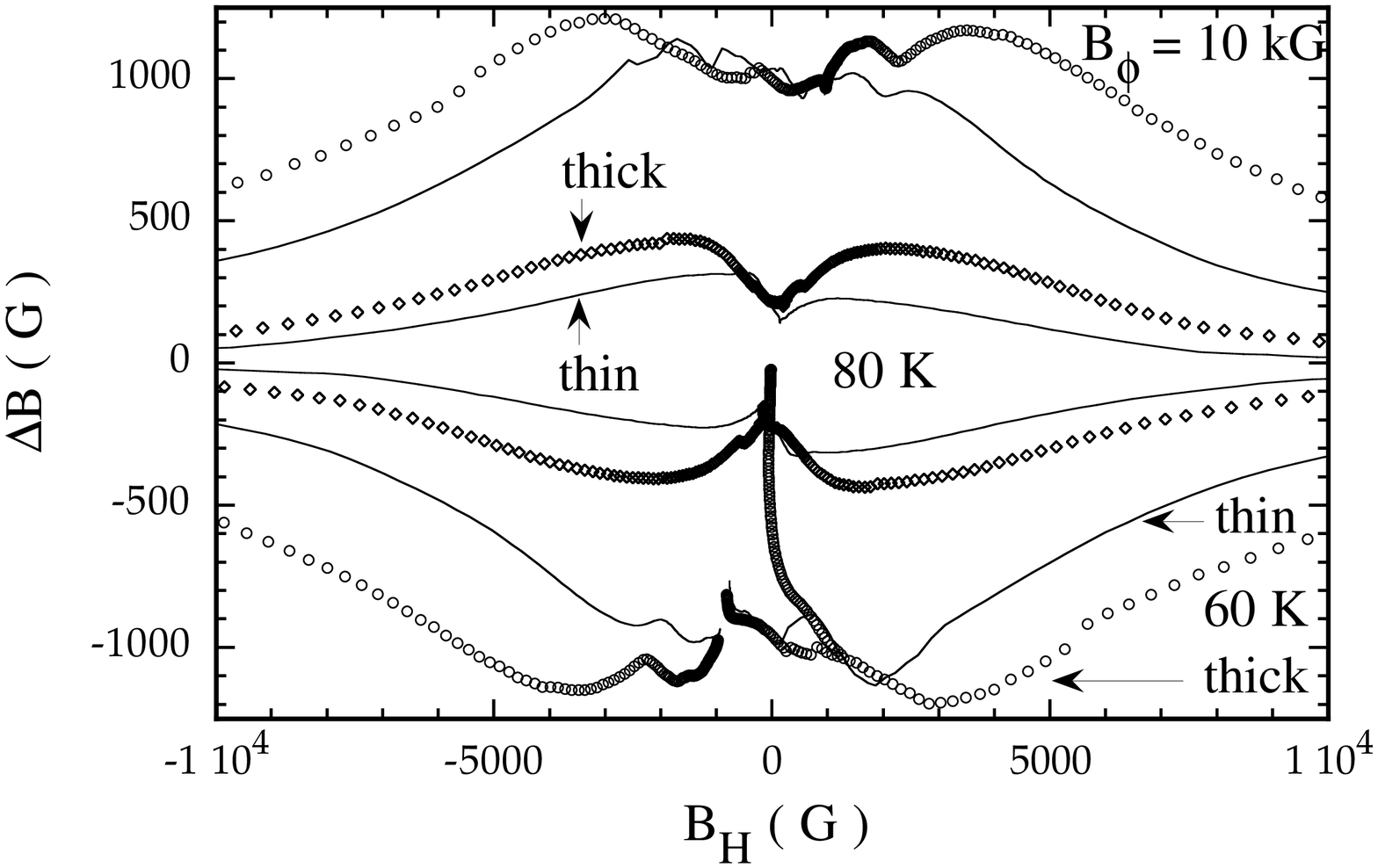}}
\caption{ \label{Loops}
Local ``magnetization'' loops $\Delta B = B_{H} - H_{a}$, measured on the
crystal shown in Fig.~\protect\ref{Half} at 60~K and 80~K.
Open symbols and thin lines represent data
taken on the thick and thin parts of the crystal, respectively. The conditions of the 
magneto-optical experiments are reproduced by the virgin 
magnetization curve and subsequent low field data, which show
a near overlap of the ``thick'' and ``thin'' data  
up to $B_{H} \approx 2$ kG for $T = 60$ K and up to 500 G for $T = 80 
K$. }
\end{figure}

Unfortunately, the magneto-optical technique is
limited to low fields, at which the self--field generated by the superconductor
is comparable to or larger than the applied field. In order to extend the
measurements
to fields comparable to $B_{\phi}$ we used the micro Hall probe
technique\cite{MicroHall}.
A small home--made single crystalline InSb Hall probe with active area
$\approx 80 \times 80$ $\mu$m$^2$ was consecutively placed in
equivalent positions
on the thick and thin parts of the crystal shown in Fig.~\ref{Half},
such that in each case its distance to the crystal ends was approximately equal to
half the crystal width. Loops of the hysteretic induction $B_{H}$  were measured 
for applied fields up to 50 kG
and temperatures 45 K $ < T < 85$ K. Fig.~\ref{Loops} shows the difference
$\Delta B = B_{H} - H_{a}$ measured on both the thick and the thin parts of
the crystal at $T = 60$ and 80 K, as function of the local value of 
$B_{H}$. The shape of these  loops is in good agreement with those in the 
literature\cite{First}.
It is seen that at low--field ($B_{H} < 2$ kG and $B_{H} < 500$ G for

\noindent  $T =60$ K 
and 80 K respectively) the``local magnetization'' $\Delta B$ measured 
on the thick and thin parts of the crystal 
{\em practically coincides} or differs by less than the amplitude of 
the low field $\Delta B$ irregularities. This contradicts the ratio 
of $\approx 2$ expected from the thickness variation for the case of 
bulk pinning,  and confirms the magneto-optical
observations. This field regime, in which vortex motion is limited by kink
nucleation at the surface, corresponds to the regime where the width
of the magnetic hysteresis loop shows a plateau ($T \lesssim 55$ K), or
increases with field ($T \gtrsim 55$ K). The loops start to deviate from
each other at the induction $B_{max}$ where $|\Delta B|$ measured on the thinner part is
maximum. A comparison with the virgin $B_{H}$--curve shows that $B_{max}$ 
is greater than the field of full flux penetration.  
Above $B_{max}$, $\Delta B$ decreases until, for
$B_{H}\gtrsim B_{\phi}$, $\Delta B$ remains constant
and displays the thickness dependence characteristic for bulk pinning.

We interpret the occurence of either surface or bulk depinning in the
different regimes of the magnetic hysteresis loop in terms of
pinning of single vortices by individual columns at low fields (each
vortex can find an empty track) and
plastic vortex creep at higher fields $\gtrsim B_{\phi}$. The
single-vortex pinning regime corresponds to that of surface depinning.
In this regime, the critical current should be estimated as $j_{c} \sim
\Delta B/ \mu_{0}\lambda$, instead of the usual $j_{c} \sim \Delta B / \mu_{0}d$. 
This yields a critical current value $j_{c} \gtrsim 10^{8}$ Acm$^{-2}$ for single
vortex depinning from a track, which at low $T$ tends to the initial estimates which had
$j_{c}$ comparable to the depairing current\cite{Brandt92III}. It is clear that
with such current values, the usual critical state in the crystal
bulk cannot exist: the self-field would generate large vortex
curvature and many ``pre-formed'' vortex kinks that would immediately
slide to the crystal equator and mutually annihilate. Thus bulk pinning can
only appear at fields when single-vortex pinning is no longer relevant.
This happens when $H_{a}$ approaches a sizeable fraction of
$B_{\phi}$: many free vortices appear in the system, as was directly observed
by scanning tunnelling microscopy\cite{FreeVortex} and revealed by model
calculations\cite{Nori}.
In this case the plastic motion of these free vortices through the ``forest''
of vortices trapped by the columnar defects determines the critical
current and the screening properties of the superconductor. Although
much lower than the current needed for depinning from a track, this
critical current is still much higher than that of the unirradiated
crystal\cite{Nelson92,Plastic}.

In conclusion, the observed thickness independence of the shielding
current in YBCO crystals with parallel columnar defects ($\parallel
c$) proves that vortex depinning from the columns occurs via surface
nucleation of vortex kinks which easily slide further down the columns
into the sample volume. The critical current is that necessary for
kink nucleation and flows only on the surface. Surface imperfections can
considerably facilitate the nucleation process; sharp surface steps can induce
a diode--like flow of vortices which should also be seen
 as an asymmetry of the magnetization loops\cite{Konczykowski97}.
Similar gigantic surface pinning may be expected for pinning by 
twin planes and for intrinsic pinning.

We gratefully acknowledge S. Bouffard for the help with the irradiation,
and Th. Schuster for discussions on the central idea of the present work.

\newpage

\end{document}